\documentclass[aps,prx,twocolumn,superscriptaddress,raggedbottom,showpacs,floatfix,longbibliography]{revtex4-1}
\usepackage{amsmath,amsfonts,amssymb,amsthm,graphics}
\usepackage[next]{inputenc}
\usepackage[colorlinks=true,citecolor=blue,linkcolor=blue]{hyperref}
\usepackage{bbm}
\usepackage{bbold}
\usepackage{booktabs}
\usepackage{multirow}
\usepackage{hhline}
\usepackage{comment}
\usepackage{float}
\usepackage{latexsym}
\usepackage[dvipsnames]{xcolor}
\usepackage{graphicx}
\usepackage{epstopdf}
\usepackage{bm}
\usepackage{verbatim}
\usepackage{changes}

\newcommand{\sech}{{\rm sech}}

\newcommand{\bs}{\boldsymbol}
\newcommand{\sgn}{{\rm sgn}}
\renewcommand{\vec}{\mathbf}

\newcommand{\x}{\chi}
\newcommand{\massterm}{m}
\newcommand{\lambdaratio}{\zeta}
\newcommand{\generalcoord}{Q}

\definecolor{TTH-color}{named}{green}
\definecolor{HH-color}{named}{magenta}
\definecolor{VG-color}{rgb}{0.97,0.57,0.11}

\definecolor{TTH-color2}{named}{red}
\definecolor{HH-color2}{named}{DarkOrchid}
\definecolor{VG-color2}{rgb}{0.87,0.47,0.01}

\begin{document}
\title{Electron Induced Massive Dynamics of Magnetic Domain Walls}
\author{Hilary M. Hurst}
\affiliation{Joint Quantum Institute, National Institute of Standards and Technology, and University of Maryland, Gaithersburg, Maryland, 20899, USA}
\affiliation{Department of Physics and Astronomy, San Jos\'{e} State University, San Jos\'{e}, California, 95192, USA}
\author{Victor Galitski}
\affiliation{Joint Quantum Institute and Condensed Matter Theory Center, Department of Physics, University of Maryland, College Park, Maryland 20742-4111, USA}
\author{Tero T.~Heikkil\"a}
\affiliation{Department of Physics and Nanoscience Center, University of Jyv\"askyl\"a, P.O. Box 35 (YFL), FI-40014 University of Jyv\"askyl\"a, Finland}

\begin{abstract}
We study the dynamics of domain walls (DWs) in a metallic, ferromagnetic nanowire. We develop a Keldysh collective coordinate technique to describe the effect of conduction electrons on rigid magnetic structures. The effective Lagrangian and Langevin equations of motion for a DW are derived. The DW dynamics is described by two collective degrees of freedom: position and tilt-angle. The coupled Langevin equations therefore involve two correlated noise sources, leading to a generalized fluctuation-dissipation theorem (FDT). The DW response kernel due to electrons contains two parts: one related to dissipation via FDT, and another `inertial' part. We prove that the latter term leads to a mass for both degrees of freedom, even though the intrinsic bare mass is zero. The electron-induced mass is present even in a clean system without pinning or specifically engineered potentials. The resulting equations of motion contain rich dynamical solutions and point toward a new way to control domain wall motion in metals via the electronic system properties. We discuss two observable consequences of the mass, hysteresis in the DW dynamics and resonant response to ac current. 

\end{abstract}

\maketitle

\section{Introduction \label{sec:introduction}}
Control of magnetic textures via electric currents is an important step toward fabricating robust magnetic memory devices~\cite{Hayashi2008, Parkin2008, Tanigawa2009, Koyama2011, Brataas2012, Catalan2012, Knoester2014}. Electrical control of magnetic domains enables devices that can be operated at low power without the high magnetic fields usually needed to induce magnetization switching~\cite{Tsoi2003, Yamanouchi2004}. The origin of electrical control in metallic ferromagnets is the interaction between current-carrying conduction electrons and domain wall (DW) magnetization~\cite{Berger1984, Brataas2012}. 

Previous theoretical and experimental work has established that in systems with hard-axis anisotropy DWs are well-described as rigid structures with two dynamical degrees of freedom, position $X$ and tilt-angle $\phi$~\cite{Slonczewski1972, Braun1996}. These `collective coordinates' are coupled due to the microscopic quantum spin dynamics, and applying external forces on $X$ or $\phi$ then leads to domain wall motion~\cite{Berger1984, Slonczewski1972, Braun1996, Tatara2004}. Already in the beginning of 1980s, Berger predicted that magnetic domain walls could be moved by application of charge currents~\cite{Berger1984}. In particular, Berger identified two ways the currents affect the domain wall motion: via direct (and non-adiabatic) forces and via adiabatic spin torques. This phenomenology was confirmed with a more microscopic approach by Tatara and Kohno who formed the now widely applied picture in terms of two dynamic coordinates~\cite{Tatara2004}. The two coordinate description can sometimes be simplified when one of the collective coordinates is `pinned' by external potentials such that the dynamical equations reduce to a single equation of motion for $X$ or $\phi$, and the relevant variable can be assigned a mass which depends on the strength of the pinning potential~\cite{Doring1948, Takagi1996, Braun1994, Duine2008}.

The existence of a domain wall mass leads to an `inertial' or delayed response to external driving via electrical currents or magnetic fields~\cite{Thomas2010}. This is particularly important for transient effects in domain wall motion, which have recently garnered considerable interest~\cite{Saitoh2004, Duine2007, Jiang2013, Sitte2016, Torrejon2016, Domenichini2019}. Domain walls can also continue moving when external driving is removed due to inertia; this allows them to be manipulated at lower power~\cite{Thomas2010}. Inertial effects appear to be system dependent and are not always observed. In some experiments essentially instantaneous DW response was observed, meaning that the DWs were effectively massless~\cite{Vogel2012}. Domain wall inertia may even be a tunable property in some materials, as found in Ref.~\cite{Torrejon2016}. Pinning and internal deformations of the domain wall have also been shown to lead to effectively massive descriptions of DWs~\cite{Koyama2011, Chauleau2010}. 

\begin{figure}[t!]
\centering
\includegraphics[scale=1.0]{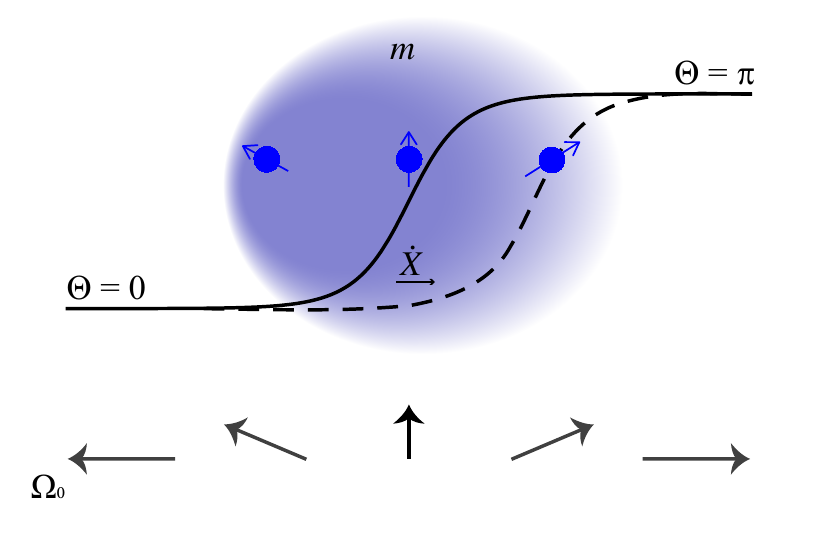}
\caption{Schematic showing DW dynamics induced by relative dynamics of electrons. The arrows on the bottom indicate local spin for a planar DW in a 1D wire with the easy axis parallel to the wire direction (horizontal). The solid black line indicates the DW angle $\Theta$ (defined in text) at $t=0$. In the adiabatic approximation, electron spins (blue arrows) align exactly with the local DW spin. When the DW is moving with velocity $\dot{X}$ relative to electrons, the electron spins do not exactly align with the profile at a later time, indicated by the dashed line. This effect results in a mass for the domain wall (shaded region). In general the electrons are not spin polarized and there are two electron `bands', one aligned with the DW and one anti-aligned. Here, only one is pictured for clarity.\label{fig:sketch} }
\end{figure}

Even though inertial effects in domain wall dynamics have been ubiquitously observed in various experiments, the fundamental origin of the domain wall mass is unclear. This is the central question we focus on in this work. By taking the dynamics of electrons into account, we show that current-driven domain walls really have an electron-induced inertial mass, independent of any system disorder or external pinning sites. Previous derivations of the effect of electrons on domain wall dynamics disregard the effect of the direct electron dynamics, i.e., the relative motion of the electrons. Here we show how taking into account these dynamics leads to additional terms in the domain wall equations of motion, in particular to massive dynamics of \emph{both} $X$ and $\phi$, shown schematically in Fig.~\ref{fig:sketch}. These additional terms have experimental consequences, and we show that DWs have resonances which can be probed via ac electric fields. Domain walls can also exhibit hysteresis in the their dynamics due to the electron-induced mass.

Our theory provides a unified framework to understand all-electrical control of domain walls and other magnetic textures. As we show in Sec.~\ref{Sec:dynamics}, the intrinsic electron fluctuations lead to domain wall motion described by a system of coupled equations
\begin{subequations}
\begin{align}
    \massterm \ddot{\phi} - \dot{\x} + \alpha\dot{\phi} + j_{\rm t} + \sin(2\phi) &= \xi_\phi(t) \label{Eqn:phi1},\\
    \massterm \ddot{\x} + \dot{\phi} + \alpha\dot{\x} &= \xi_\x(t) \label{Eqn:x1}, 
\end{align} 
\label{Eqn:eomall}
\end{subequations}

\noindent \noindent where $\x = X/\lambda$ is the dimensionless position of the DW, $\alpha$ is a damping parameter, and $j_{\rm t}$ is the spin torque. The noises $\xi_\x, \xi_\phi$ are Langevin stochastic variables whose correlation functions are discussed in Sec.~\ref{Sec:dynamics}. Interestingly, these dynamics give rise to multicomponent noise which requires a nontrivial generalization of the fluctuation-dissipation theorem (FDT).

The dimensionless DW mass due to electrons in a ferromagnet is 
\begin{equation}
\massterm = \frac{K_\perp}{2\Delta}\frac{s}{N}.
\label{Eqn:mass}
\end{equation}
The time scale is set in units of the anisotropy, $t \rightarrow K_\perp t/2\hbar$ where $\hbar$ is Planck's constant.  Here $K_\perp$ is the hard-axis anisotropy energy of the magnet, $N$ is the number of localized spins in the domain wall, $\Delta$ is the strength of exchange coupling between the DW and conduction electrons, and $s=(k_{F\downarrow}-k_{F\uparrow})\lambda/(2 \pi)$ is the amount of electron spin within the DW. The Fermi momenta of the electrons with spin aligned ($\uparrow$) or anti-aligned ($\downarrow$) with the local domain wall spin is $k_{\rm F \uparrow(\downarrow)}$.

We study an electronic system with quadratic dispersion, giving $k_{F\downarrow/\uparrow} = \sqrt{2 m_{\rm e}(\mu \pm \Delta)/\hbar^2}$ where $m_{\rm e}$ is the electron mass and the spin-dependent Fermi levels are $\mu \pm \Delta$ for electron chemical potential $\mu$. In this case, for an unpolarised electron system such that $\Delta < \mu$, $m \approx K_\perp \lambda /N h v_{\rm F}$, where $h = 2\pi\hbar$ and $v_{\rm F}$ is the Fermi velocity. The mass of the domain wall is therefore dependent on the time $\tau_{\rm e} \sim \lambda/v_{\rm F}$ it takes the electron to traverse the DW width. The faster electrons travel through the domain wall region, the smaller the effective mass. We emphasize that the mass is dependent on both the magnetic system properties and the properties of the electronic system. In typical metallic systems $m$ is small because $v_{\rm F}$ is large, however in systems with large $K_\perp$ or small $v_{\rm F}$ it becomes relevant.

Our general theory opens the door to manipulating the inertial properties of domain walls by tuning the properties of the electronic system. The formalism could also be adapted to study effects in semiconductors or disorderd systems, and may explain why inertial phenomena appears to be system dependent. Using Eqs.~\eqref{Eqn:eomall} to model domain wall dynamics leads to a host of rich dynamical phenomena that can be experimentally probed. Furthermore, using these equations one can establish a fruitful analogy between DW dynamics and the dynamics of Josephson junctions, using known results of the latter to gain intuition about DWs~\cite{Braun1994, barone1982physics, terosbook}.

The paper is structured as follows: In Sec.~\ref{Sec:model} we define the model and use the Keldysh formalism to describe the influence of conduction electrons on the motion of magnetic textures, such as domain walls. In Sec.~\ref{Sec:dynamics} we derive the general equations of motion for the domain wall and prove a generalized fluctuation-dissipation theorem (FDT). We show that the response kernel contains two parts, one responsible for dissipation in the usual sense (related to the correlations of noise) and another part, which we term `inertial', is responsible for the mass. These first two technical sections are devoted to the general formalism, which can be applied to other systems.

In the remaining sections we discuss the specific case of a wide domain wall and implications for experiments. In Sec.~\ref{Sec:Adiabatic} we calculate the response kernel for DW motion due to electrons and in Sec.~\ref{Sec:Lagrangian} we show that this leads to a new effective DW Lagrangian with additional terms. In Sec.~\ref{Sec:Experiment} we investigate two possible signatures of the mass; resonant excitation of the DW and hysteresis in the DW dynamics. We also discuss how the hysteresis problem is related to well-known dynamics of Josephson junctions. Finally, in Sec.~\ref{Sec:Discussion} we discuss future questions and directions where our formalism may be insightful. Additional technical details of the calculations are included in Appendices~\ref{App:response}-\ref{App:F}. 

\section{Model and Assumptions \label{Sec:model}}
We consider a model system of a quasi one-dimensional metallic ferromagnet with localized spins $S$ and free conduction electrons. The total system is described by an action $\mathcal{S} = \mathcal{S}_{\rm m} + \mathcal{S}_{\rm e}$, where $\mathcal{S}_{\rm m}$ denotes the action of the magnetic moments and $\mathcal{S}_{\rm e}$ describes conduction electrons including their coupling to the magnetization.

\subsection{Planar Domain Wall}
The magnetization profile is described by the continuum action 
\begin{equation}
    \mathcal{S}_{\rm m} = \int dt \frac{dx}{a}~\hbar S\vec{A}\left[\bs{\Omega}\right]\cdot\dot{\bs{\Omega}}-\hat{H}_{\rm m}\left[\bs{\Omega}\right], 
\end{equation}
where $a$ is the system lattice constant, $\bs{\Omega}(x) = (\sin\Theta(x)\cos\Phi(x), \sin\Theta(x)\sin\Phi(x), \cos\Theta(x))$ is a three-dimensional unit vector parameterizing the direction of magnetization, and $\nabla_{\bs{\Omega}} \times \vec{A} = \bs{\Omega}$ is the effective vector potential accounting for the quantum spin dynamics~\cite{Braun1996, Altland2010}. The magnetic Hamiltonian is 
\begin{equation}
   H_{\rm m}\left[\bs{\Omega}\right] = \frac{S^2}{2}\int \frac{dx}{a}~J(\nabla\bs{\Omega})^2 -K_z\Omega_z^2 + K_\perp\Omega_y^2 ,
   \label{Eqn:Hmag}
\end{equation}
where $J, K_z,$ and $K_\perp$ are positive coefficients for the spin stiffness, easy-axis anisotropy, and hard-axis anisotropy respectively. Hamiltonian~\eqref{Eqn:Hmag} has a classical planar domain wall solution~\cite{Schryer1974}
\begin{equation}
    \Theta = 2\arctan\left[\exp\left(\frac{X-x}{\lambda}\right)\right]\mbox{~~;~~}\Phi = \phi,
    \label{Eqn:DWsol}
\end{equation}
\noindent where $\lambda = \sqrt{J/K_z}$ is the domain wall width, $X$ is the domain wall position and $\phi$ is a constant. We consider a N\'{e}el wall with $\phi = 0$, which occurs in systems with hard-axis anisotropy~\cite{Tatara2008, Emori2013}. In order to study domain wall dynamics, $X$ and $\phi$ are promoted to dynamical quantities $X(t)$ and $\phi(t)$ ; these are the collective coordinates of the domain wall~\cite{Takagi1996, Tatara2004}. 

The collective coordinate description assumes that the domain wall is rigid, without deformation such that $X$ and $\phi$ are the only dynamical coordinates, resulting from the zero energy spin wave modes~\cite{Tatara2008}. Spin wave modes describing wall deformation have an energy gap $\sim\sqrt{K_z K_\perp}$ which we consider to be large compared to the other energy scales in this problem.

Integrating over the spatial degrees of freedom results in the action for the collective coordinates
\begin{equation}
    \mathcal{S}_{\rm dw}\left[X,\phi\right] = NS\int dt~\hbar\dot{\x}\phi - \frac{K_\perp S}{2}\sin^2(\phi)
    \label{Eqn:SDW}
\end{equation}
where $\dot{\x} = \dot{X}/\lambda$ and $N=2\lambda/a$ is the number of spins in the domain wall. From the term $\propto \dot{\x}\phi$ is it is clear that $\x$ and $\phi$ are intrinsically coupled via a gauge-like term, which has important consequence for the dynamics~\cite{Braun1996}. In the following we set $S=1$, and $\dot{\x}$ and $\dot{\phi}$ both have units of time$^{-1}$.

\subsection{Conduction Electrons \label{SubSec:ElectronsL}}
Conduction electrons couple to the magnetization field $\bs{\Omega}(x)$ via a local exchange interaction. The action for the electrons is 
\begin{equation}
    \mathcal{S}_{\rm e} = \int dtdt'\int dx~\bar{\tilde{\bs{\psi}}}(x, t) \left\{i\hbar\partial_{t'} - H\left[\bs{\Omega}(x,t)\right]\right\}\tilde{\bs{\psi}}(x,t').
    \label{Eqn:Electronaction1}
\end{equation}
where 
\begin{equation}
    H\left[\bs{\Omega}\right] = \left(-\frac{\hbar^2\nabla^2}{2m_{\rm e}}-\mu\right)\hat{\tau}^0 -\Delta \bs{\Omega}(x,t)\cdot\hat{\bs{\tau}}
    \label{Eqn:ElectronHamiltonian1}
\end{equation}
is the Hamiltonian for electrons including the spatially varying magnetization profile and chemical potential $\mu$, and $\hat{\tau}^0$ is the $2\times2$ identity matrix. We use the notation $\hat{\cdot}$ to denote a matrix in the electron spin space. The exchange coupling strength is $\Delta > 0$ and $\hat{\bs{\tau}} = (\hat{\tau}^1, \hat{\tau}^2, \hat{\tau}^3)$ is the vector of Pauli matrices acting on electron spin. Equation~\eqref{Eqn:Electronaction1} is written in terms of Grassmann spinors $\bar{\tilde{\bs{\psi}}}(x,t)$ and $\tilde{\bs{\psi}}(x,t)$,  
where $\tilde{\bs{\psi}} = (\tilde{\psi}_\uparrow, \tilde{\psi}_\downarrow)^T$ is a two-component spinor and likewise for $\bar{\tilde{\bs{\psi}}}$. 

The electron action in Eq.~\eqref{Eqn:Electronaction1} presents a theoretical challenge because the exchange interaction $\Delta\bs{\Omega}(x,t)$ varies in space and time. Previous works used a local gauge transformation to diagonalize the exchange interaction, transferring information about the domain wall dynamics to a fluctuating gauge field (see~\cite{Tatara2008} for a review).

However, if the domain wall is treated as a rigid object then the description of the DW dynamics is reduced to only two variables, $\x(t)$ and $\phi(t)$. In this case electrons couple separately to the static domain wall $\bs{\Omega}_0(x)$ and the dynamical coordinates, as we show below. This method is inspired by similar treatment of topological defects in other systems~\cite{Rajaraman1987, Kovrizhin2001, Efimkin2016, Psaroudaki2017, Kim2018}.

We can write $\bs{\Omega}(x)$ as a function of the collective coordinates, 
$\bs{\Omega}(x-\lambda \x(t), \phi(t))$ and redefine the electron fields via the transformation 
\begin{subequations}
\begin{align}
\tilde{\bs{\psi}}(x, t) &= \exp\left[i\frac{\hat{\tau}^3\phi(t)}{2}\right]\bs{\psi}(x-\lambda\x(t),t),\\
\bar{\tilde{\bs{\psi}}}(x, t) &=\exp\left[-i\frac{\hat{\tau}^3\phi(t)}{2}\right]\bar{\bs{\psi}}(x-\lambda\x(t),t).
\label{Eqn:Transform}
\end{align}
\end{subequations}
The new action for the electrons is
\begin{align} 
    \mathcal{S}_{\rm e} &= \int dtdt' \int dx~\bar{\bs{\psi}}(x, t) \left[i\hbar\partial_{t'} -H\left[\bs{\Omega}_0(x)\right]\right]\bs{\psi}(x,t')\nonumber \\
    &- \int dtdt'\int dx~\bar{\bs{\psi}}(x,t)\left[i\hbar\lambda\dot{\x}\partial_x + \frac{\hbar}{2}\dot{\phi}\hat{\tau}^3\right]\bs{\psi}(x,t'),
    \label{Eqn:ElectronAction}
\end{align}
where $H\left[\bs{\Omega}_0(x)\right] = -\hbar^2\partial^2_x/2m_{\rm e} -\Delta \bs{\Omega}_0(x)\cdot\hat{\bs{\tau}} - \mu$ is now a \emph{time-independent} Hamiltonian that describes free electrons coupled to a static, rigid domain wall $\bs{\Omega}_0(x)$. The second and third terms of Eq.~\eqref{Eqn:ElectronAction} directly couple the electrons to $\dot{\x}(t)$ and $\dot{\phi}(t)$. Thus, when the domain wall is in motion relative to the electronic `bath', these additional terms affect the dynamics of the electrons which in turn has consequences for domain wall motion. 

Equation \eqref{Eqn:ElectronAction} constitutes the starting point for our treatment of conduction electrons. The strategy is as follows: first, we define a diagonal basis for the electrons in the presence of a static domain wall. This allows us to treat the term with $H[\bs{\Omega}_0(x)]$ exactly. Then, we treat fluctuations around the static solution as a perturbation. In the limit of a slowly-moving DW we can integrate out the electrons and find an effective description of the DW dynamics.

We first consider the consequences of this procedure in the general case, then present exact results for a wide domain wall varying adiabatically compared to the electron Fermi wavelength in Sec.~\ref{Sec:Adiabatic}. Our formalism is not restricted only to the adiabatic case. 

For any static domain wall profile $\bs{\Omega}_0(x)$ there exists a set of `domain wall basis' functions $\lbrace\bs{\varphi}_{\sigma k}(x)\rbrace$ for electrons. The basis functions are two component spinors such that $H\left[\bs{\Omega}_0(x)\right]\bs{\varphi}_{\sigma k}(x) = \varepsilon_{\sigma k}\bs{\varphi}_{\sigma k}(x)$, where $\varepsilon_{\sigma k}$ is the energy. The indices $k$ and $\sigma$ label single-particle eigenstates of the Hamiltonian. Here, $k$ is a momentum-like variable and $\sigma$ labels electron `bands' whose spin is everywhere aligned or anti-aligned with the local domain wall spin. Even though the DW breaks translational symmetry, we can still define eigenstates in terms of $k$ and $\sigma$ provided we find appropriate basis functions; this is well established in soliton theory~\cite{Kovrizhin2001, Efimkin2016}. In Sec.~\ref{Sec:Adiabatic} we present a specific case where  $\lbrace\bs{\varphi}_{\sigma k}(x)\rbrace$ is calculated analytically. 

Using this basis, the electron Grassmann fields are 
\begin{subequations}
\begin{align}
    \bs{\psi}(x,t) &= \sum_{\sigma k} \bs{\varphi}_{\sigma k}(x)c_{\sigma k}(t)\\
    \bar{\bs{\psi}}(x,t) &= \sum_{\sigma k} \bs{\varphi}_{\sigma k}^*(x)\bar{c}_{\sigma k}(t),
\end{align} 
\end{subequations}
where $\bar{c}_{\sigma k}(t)$, $c_{\sigma k}(t)$ are time-dependent Grassmann numbers. The electron action now takes the form 
\begin{align}
\mathcal{S}_{\rm e} = \int dtdt'\sum_{\substack{\sigma\sigma'\\kk'}}~&\bar{c}_{\sigma k}(t)\left[i\hbar\partial_{t'} -\varepsilon_{\sigma k}\right]c_{\sigma k}(t')\nonumber \\
&-\dot{\generalcoord}^i(t)~\bar{c}_{\sigma k}(t)~^i\!V^{\sigma\sigma'}_{kk'}c_{\sigma'k'}(t). 
\label{Eqn:FinalAction}
\end{align}
\noindent Here we introduce a compact notation $\dot{\generalcoord}^i(t)$ for the generalized collective coordinates. We use latin indices $i, j$  to denote the coordinates $\x, \phi$ (e.g. $\dot{\generalcoord}^\x = \dot{\x}$) and repeated indices are summed over. Equation \eqref{Eqn:FinalAction} is convenient to work with because the first term is diagonal in $\sigma, k$ space and it is the DW \emph{dynamics} $\dot{\generalcoord}$ which perturb the electrons. Written in this form the theory lends itself to a perturbative analysis in the regime of a slow domain wall where $\dot{\x} \ll v_{\rm F}/\lambda$ and $\dot{\phi} \ll \Delta$. 

The matrix elements $^i\!V^{\sigma\sigma'}_{kk'}$ mediate scattering between domain wall basis states $|\sigma k\rangle$ and $|\sigma' k'\rangle$ due to domain wall motion, with
\begin{subequations}
\begin{align}
^\x\!V^{\sigma\sigma'}_{kk'} &= \frac{i\hbar}{\lambdaratio} \int dx~\bs{\varphi}^\dagger_{\sigma k}(x)\partial_x\bs{\varphi}_{\sigma' k'}(x),\\
^\phi\!V^{\sigma\sigma'}_{kk'} &= \frac{\hbar}{2} \int dx~ \bs{\varphi}^\dagger_{\sigma k}(x)\hat{\tau}^3\bs{\varphi}_{\sigma' k'}(x). 
\end{align}
\end{subequations}
where $\lambdaratio = \lambda_{\rm F}/\lambda$ and we have selected $\lambda_{\rm F}$ as the unit scale, making $x, k$ dimensionless.

\subsection{Keldysh Action}
Combining Eqns.~\eqref{Eqn:SDW} and \eqref{Eqn:FinalAction}, we follow the usual procedure to derive the Keldysh action defined on the contour $\mathcal{C}$,
\begin{equation}
    \int_{\mathcal{C}} dt~\mathcal{L}^{\rm K}(t) = \int^\infty_{-\infty} dt~\mathcal{L}^+(t) + \int^{-\infty}_{\infty} dt~ \mathcal{L}^-(t). 
\end{equation}
Where $\pm$ denotes the upper and lower branches of the contour, respectively~\cite{Altland2010, Kamenev2011}. The original system has four degrees of freedom: domain wall coordinates $\vec{\generalcoord}(t) = \left(\x(t), \phi(t)\right)$ and electron Grassmann variables $\bar{c}_{\sigma k}(t), c_{\sigma k}(t)$, which are re-written $c\rightarrow c^\pm$, $\vec{\generalcoord} \rightarrow \vec{\generalcoord}^{\pm}$. We then perform the Keldysh rotation via the variable transformation 
\begin{equation}
    \generalcoord^{i\pm}(t) = \generalcoord^i_{\rm c}(t) \pm \frac{\generalcoord^i_{\rm q}(t)}{2}; 
\end{equation}
subscripts c and q denote the `classical' and `quantum' parts. The Grassmann numbers transform as~\cite{Kamenev2011}
\begin{equation}
    c^{\pm}_{\sigma k} = \frac{c^1_{\sigma k} \pm c^2_{\sigma k}}{\sqrt{2}} \mbox{~~;~~} \bar{c}^{\pm}_{\sigma k} = \frac{\bar{c}^2_{\sigma k} \pm \bar{c}^1_{\sigma k}}{\sqrt{2}}.
\end{equation}
The Keldysh action for the full system is $\mathcal{S}^{\rm K} = \mathcal{S}^{\rm K}_{\rm dw} + \mathcal{S}^{\rm K}_{\rm e}$ where 

\begin{align}
\mathcal{S}^{\rm K}_{\rm dw} = \int dt&~ \hbar N\left[\dot{\generalcoord}^\x_{{\rm c}}(t)\generalcoord^\phi_{{\rm q}}(t) -\dot{\generalcoord}^\phi_{{\rm c}}(t)\generalcoord^\x_{{\rm q}}(t)\right]\nonumber \\
&-\frac{K_\perp N}{2}\sin\left[2\generalcoord^\phi_{{\rm c}}(t)\right]\generalcoord^\phi_{{\rm q}}(t),\label{Eqn:Kdw} \\
\mathcal{S}^{\rm K}_{\rm e} = \int dtdt'&  \sum_{\substack{\sigma\sigma'\\k k'}} ~\left\lbrace\vec{\bar{c}}_{\sigma k}(t)\check{G}^{-1}_{\sigma k}(t,t')\vec{c}_{\sigma k}(t')\right.\nonumber \\
&\left.-^i\!V^{\sigma\sigma'}_{kk'}\vec{\bar{c}}_{\sigma k}(t)\check{\generalcoord}^i(t)\vec{c}_{\sigma'k'}(t)\right\rbrace.\label{Eqn:Kelectron}
\end{align}

Here we use the $\check{\cdot}$ notation for matrices in Keldysh space. The matrix $\check{G}^{-1}_{\sigma k}(t,t')$ denotes the electronic Green function matrix. The Keldysh-space vectors $\vec{c} = (c^1, c^2)^T$, $\vec{\bar{c}} = (\bar{c}^1, \bar{c}^2)^T$ are coupled to the collective coordinates via the matrix
\begin{equation}
    \check{\generalcoord}^i(t) = \dot{\generalcoord}^i_{{\rm c}}(t)\check{\tau}^0 + \frac{\dot{\generalcoord}^i_{{\rm q}}(t)}{2}\check{\tau}^1.
\end{equation}
The action Eq.~\eqref{Eqn:Kelectron} is quadratic in the Grassmann fields and the electrons can therefore be integrated out. To one loop order this gives an effective action for the domain wall $\mathcal{S}^{K} \approx \mathcal{S}^K_{\rm dw} + \mathcal{S}'$ with

\begin{align}
\mathcal{S}' &= \int dt~F^i \generalcoord^i_{{\rm q}}(t) -\int dtdt'~\generalcoord^i_{\rm q}(t)\eta^{ij}(t-t')\dot{\generalcoord}^j_{\rm c}(t)\nonumber\\
&+\frac{i}{2}\int dtdt'~\generalcoord^i_{\rm q}(t)\mathcal{C}^{ij}(t-t')\generalcoord^j_{\rm q}(t').
\label{Eqn:EffectiveS}
\end{align}

The first term in Eq.~\eqref{Eqn:EffectiveS} describes how non-equilibrium forces $F^i$ act on the domain wall; these terms give the familiar spin transfer ($F^\phi$) and momentum transfer ($F^{\x}$) forces which are known to affect domain wall motion out of equilibrium~\cite{Tatara2004}. The second term contains the response kernel $\eta^{ij}(t-t')$, which in general is nonlocal in time and leads to both dissipation and mass renormalization~\cite{Altland2010, Caldeira1985}.

Finally, the third term in Eq.~\eqref{Eqn:EffectiveS} is quadratic in $\generalcoord^i_{{\rm q}}(t)$ and describes quantum fluctuations in domain wall motion. The quantum terms can be decoupled in Eq.~\eqref{Eqn:EffectiveS} by introducing a vector of auxiliary noise fields $\xi^i(t)$ using a standard Hubbard-Stratonovich transformation~\cite{Altland2010, Kamenev2011}.

\section{Domain Wall Dynamics and Fluctuation-Dissipation Theorem\label{Sec:dynamics}}

Minimizing $\mathcal{S}^{\rm K}$ with respect to $\generalcoord^i_q(t)$ leads to coupled Langevin equations of motion for $\dot{Q}$,
\begin{widetext}
\begin{subequations}
\begin{align}
    \hbar N\dot{\phi} + F^{\x} + \int dt'\eta^{\x i}(t-t')\dot{\generalcoord}^i(t') &= \xi_{\x}(t), \label{Eqn:phieom}\\
    -\hbar N\dot{\x} + F^{\phi} + \frac{K_\perp N}{2}\sin(2\phi) + \int dt'\eta^{\phi i}(t-t')\dot{\generalcoord}^i(t') &= \xi_{\phi}(t) \label{Eqn:Xeom}.
\end{align}
\label{Eqn:dwdynamicsgeneral}
\end{subequations}
\end{widetext}
The noise is characterized by the correlation function
\begin{equation}
    \langle \xi^i(t)\xi^j(t')\rangle = \mathcal{C}^{ij}(t-t'). 
    \label{Eqn:Correlator}
\end{equation}
\noindent We drop the c subscript because all further descriptions of the dynamics are in terms of classical quantities. We first consider the case $F^{i} = 0$ for electrons in equilibrium; we discuss the finite spin-torque case in Sec.~\ref{Sec:Experiment}. The kernel $\eta^{ij}(t-t')$ and correlator $\mathcal{C}^{ij}(t-t')$ are related via the fluctuation-dissipation theorem. This can be seen from the Fourier space representation, where we can write
\begin{align}
    \eta^{ij}(\omega) &= \frac{1}{\omega}\left[J^{ij}(\omega)+if^{ij}(\omega)\right]\label{Eqn:etaomega}\\
    \mathcal{C}^{ij}(\omega) &= \coth\left(\frac{\hbar\omega}{2T}\right)J^{ij}(\omega), \label{Eqn:Comega}
\end{align}
with
\begin{widetext}
\begin{align}
    J^{ij}(\omega) &= \frac{\pi\hbar^2}{2}\sum_{\substack{\sigma\sigma' \\k k'}}~^i\!V^{\sigma\sigma'}_{kk'}~^j\!V^{\sigma'\sigma}_{k'k}\left[h_{\sigma'k'}-h_{\sigma k}\right] (\varepsilon_{\sigma' k'}-\varepsilon_{\sigma k})^2\delta\left[\hbar\omega-(\varepsilon_{\sigma' k'}-\varepsilon_{\sigma k})\right]\label{Eqn:dissipative},\\
    f^{ij}(\omega) &= \frac{\hbar^2\omega^2}{2}\sum_{\substack{\sigma\sigma'\\k k'}}~^i\!V^{\sigma\sigma'}_{kk'}~^j\!V^{\sigma'\sigma}_{k'k}\left[\frac{h_{\sigma'k'}-h_{\sigma k}}{\hbar\omega - (\varepsilon_{\sigma' k'}-\varepsilon_{\sigma k})}\right],\label{Eqn:inertial}
\end{align}
\end{widetext}
where $h_{\sigma k} = \tanh[(\varepsilon_{\sigma k}-\mu)/2T]$. The spectral function $J^{ij}(\omega)$ describes the dissipative part and $f^{ij}(\omega)$ is an 
`inertial' part. Additional details on how to derive these expressions are provided in Appendix~\ref{App:response}.

Equations~\eqref{Eqn:Correlator} - \eqref{Eqn:inertial} constitute the main result of the formalism we developed, and are a generalization of the FDT. The dynamics of $\x$ and $\phi$ are coupled via a matrix response kernel $\eta^{ij}(\omega)$, which alters dynamics of the system. These expressions are not restricted to a particular description of the electrons. Furthermore, the noise correlation function $C^{ij}(t-t')$ in Eq.~\eqref{Eqn:Correlator} is not generally diagonal in $i,j$, which leads to correlated noise in different channels.
Matrix dissipation naturally arises in this problem but has not been previously discussed for magnetic DWs. We emphasize that \emph{both} the dissipative and inertial parts of $\eta^{ij}(\omega)$ can contribute to domain wall dynamics.

\section{Response Kernel for an Adiabatic Domain Wall \label{Sec:Adiabatic}}
Spin textures in ferromagnetic systems are often slowly varying in comparison to the electron length scale and therefore the adiabatic approximation is justified. Using the classical planar DW solution in Eq.~\eqref{Eqn:DWsol}, the DW forms a spin-dependent potential for electrons 
\begin{equation}
  \Delta\bs{\Omega}_0(\lambdaratio x)\cdot\hat{\bs{\tau}} = \Delta\tanh\left(\lambdaratio x\right)\hat{\tau}^1 + \Delta\sech\left(\lambdaratio x \right)\hat{\tau}^3,  
  \label{Eqn:DWpotential}
\end{equation}
where $\lambdaratio = \lambda_{\rm F}/\lambda$. If the domain wall is wide enough that $\lambdaratio \ll 1$, this potential is slowly varying in space compared to the Fermi wavelength of the electrons. We assume that the electron spin adiabatically follows the spin of the static domain wall. The electron Hamiltonian can then be treated using WKB methods~\cite{Littlejohn1991}. The eigenstates for the potential are 
\begin{subequations}
\begin{align}
    \bs{\varphi}_{\uparrow k}(\lambdaratio x) &= \frac{1}{\sqrt{1+e^{-2\lambdaratio x}}}
    \begin{pmatrix}
    -e^{-\lambdaratio x} \\
    1
    \end{pmatrix}
    e^{ikx}
    \label{Eqn:psiup}
    \\
    \bs{\varphi}_{\downarrow k}(\lambdaratio x) &= \frac{1}{\sqrt{1+e^{-2\lambdaratio x}}}
    \begin{pmatrix}
    1\\
    ~~e^{-\lambdaratio x}
    \end{pmatrix}
    e^{ikx},
    \label{Eqn:psidown}
\end{align}
\label{Eqn:botheigenstates}
\end{subequations}
\noindent where $x, k$ are dimensionless. Details of the calculation of $\bs{\varphi}_{\sigma k}(\lambdaratio x)$ are provided in Appendix~\ref{App:eigenstates}. Equations~\eqref{Eqn:botheigenstates} form an orthonormal set and a complete basis for the electrons with $\varepsilon_{\uparrow/\downarrow,k} = \hbar^2k^2/2m\lambda_{\rm F}^2 \pm \Delta - \mu$. Therefore, we can apply the general formalism developed in Secs.~\ref{Sec:model}-\ref{Sec:dynamics}. 

Here we summarize the results for the response function $\eta^{ij}(\omega)$ and present more detailed calculations in Appendix~\ref{App:F}. Recall from Sec.~\ref{SubSec:ElectronsL} that domain wall motion mediates scattering between the domain wall basis states $|\sigma k\rangle$ and $|\sigma'k'\rangle$. In the adiabatic approximation, we find that intraband scattering ($\sigma = \sigma'$) is exactly zero in all cases. The movement of the domain wall therefore only mediates scattering between the bands, i.e., $\sigma \neq \sigma'$. Since the bands here are the relative to the local DW spin, $\bs{\varphi}_{\uparrow k}$ denotes eigenstates where the electron spin is everywhere aligned with the local magnetization $\bs{\Omega}_0(x)$, and for $\bs{\varphi}_{\downarrow k}$ the electron spin is anti-aligned with $\bs{\Omega}_0(x)$. 

At frequencies below the electronic gap, $\hbar\omega \lesssim 2\Delta$, we find that the inertial term $f^{ij}(\omega)$ is the only relevant one, with diagonal and off-diagonal terms
\begin{subequations}
\begin{align}
    f^{ii}(\omega) &\approx \frac{4\Delta\hbar^2\omega^2s}{(\hbar\omega)^2-4\Delta^2}\label{Eqn:fdiag}\\
    f^{\phi \x}(\omega) &\approx \frac{2i\hbar^3\omega^3s}{(\hbar\omega)^2-4\Delta^2},\label{Eqn:foffdiag}
\end{align}
\label{Eqns:fall}
\end{subequations}
\noindent where $f^{\x\phi} = -f^{\phi \x}$. The parameter $s = (k_{\rm F \downarrow} - k_{\rm F \uparrow})\lambda/2\pi$ is the amount of electron spin within the domain wall width $\lambda$. Here we assume a quadratic dispersion relation, with $\Delta < \mu$ and $k_{F\downarrow/\uparrow} = \sqrt{2 m_{\rm e}(\mu \pm \Delta)/\hbar^2}$. At low frequencies $\hbar\omega \ll 2\Delta$, we have $f^{\phi \x} \sim \mathcal{O}(\omega^3)$ and $f^{ii}(\omega) = -\hbar^2\omega^2s/\Delta$. The diagonal part is therefore dominant and leads to a mass  $M_{\rm dw} = \hbar^2s/\Delta$ in the equations of motion.

The low-frequency ($\hbar\omega < 2\Delta$) contribution of the spectral function $J^{ij}(\omega)$ is zero in this case. Like the inertial part, we find that $J^{ij}(\omega)$ is exactly zero for intraband scattering ($\sigma = \sigma'$). The contribution from interband scattering, which requires a spin flip of the electron, is only nonzero for frequencies at the electronic gap $\hbar\omega \approx \pm 2\Delta$ since $J^{ij}(\omega)\propto\delta(\hbar\omega \pm 2\Delta)$. This is beyond the low frequency approximation we consider below.

Therefore, this theory does not describe damping or dissipation due to electrons. An additional mechanism of spin or momentum relaxation, such as that caused by disorder, must be included. It is not sufficient simply to include a finite electron lifetime in the equilibrium Green's functions, in this case we have checked that Ohmic friction (damping) is still exactly zero. This result is consistent with previous theories of conduction electrons adiabatically interacting with a domain wall spin, where dissipation does not arise~\cite{Tatara2008}. It is well established that Gilbert damping and other dissipation mechanisms are ubiquitous in solid state systems, and incorporating additional dissipative effects into this theory will be the subject of future work. 

\section{Dynamics in the Adiabatic Approximation \label{Sec:Lagrangian}}

We now consider the dynamics of the DW in the low-frequency limit. Inserting the response kernel Eqs.~(\ref{Eqn:etaomega},\ref{Eqns:fall}) to the equations of motion Eqs.~(\ref{Eqn:phieom}-\ref{Eqn:Xeom}) and assuming $\hbar\omega \ll 2\Delta$ leads to 

\begin{subequations}
\begin{align}
M_{\rm dw} \ddot \phi- \hbar N  \dot \x + \frac{N K_\perp}{2} \sin(2\phi) + F^\phi &= \xi_\phi(t)\\
M_{\rm dw} \ddot \x + \hbar N  \dot \phi + F^\x &= \xi_\x(t).
\end{align}
\label{Eqn:dwdynamics1}
\end{subequations}
\noindent The second derivative terms result from the dominant low-frequency contribution of the response kernel. This term describes the inertial effect of conduction electrons on the domain wall. It can hence be interpreted as a ``mass'' $M_{\rm dw}$ of the domain wall.

Equations \eqref{Eqn:dwdynamics1} present the domain wall dynamics as a result of a system with two coupled coordinates. Namely, this equation of motion can be obtained from the Lagrangian
\begin{equation}
\mathcal{L}_{\rm dw} = \frac{M_{\rm dw}}{2} (\dot \phi^2 + \dot \x^2)+ \hbar N\dot \x \phi - V(\x,\phi),
\label{eq:dwlagrangian}
\end{equation}
where 
\begin{equation}
    V(\x,\phi)=-\frac{N K_\perp}{4} \cos(2\phi)+F^\phi \phi + F^\x \x
\end{equation}
is the effective potential. Note that the second term in Eq.~\eqref{eq:dwlagrangian} could also be written as $-\hbar N \x \dot \phi$, as it produces the same dynamics. On the level of the action, these two terms can be obtained from each other via partial integration. From $\mathcal{L}_{\rm dw}$ the conjugate momenta are
\begin{equation}
p_\phi = M_{\rm dw} \dot \phi \mbox{~~;~~} p_{\x} = M_{\rm dw} \dot \x + \hbar N\phi, 
\end{equation}
so that the effective DW Hamiltonian is
\begin{equation}
H_{\rm dw} = \frac{p_\phi^2}{2M_{\rm dw}} + \frac{(p_{\x}- \hbar N\phi)^2}{2M_{\rm dw}} + V(\x,\phi).
\label{Eqn:dwhamiltonian}
\end{equation}
The coupling between the coordinates is thus similar to the gauge field coupling in electrodynamics.

Here we consider further the consequences of the electron-induced mass. To do this, we make Eq.~\eqref{Eqn:dwdynamics1} dimensionless by introducing a dimensionless time scale $t'=K_\perp t/2\hbar$. We use the parameter $\massterm = K_\perp s/2\Delta N$ defined in Sec.~\ref{sec:introduction} to describe the relative importance of the mass. We also define the dimensionless spin torque $j_{\rm t}=2 F^\phi/(N K_\perp)$ and force $f_\x=2 F^\x/(N K_\perp)$ terms. With these definitions, the domain wall equations of motion become
\begin{subequations}
\begin{align}
\massterm \ddot \phi - \dot \x + \alpha \dot \phi + j_{\rm t} + \sin(2\phi) &= \tilde{\xi}_\phi(t'),\label{Eqn:dwdynamicsphi}\\
\massterm \ddot \x + \dot \phi + \alpha \dot \x + f_x &= \tilde{\xi}_\x(t').
\end{align}
\label{Eqn:dwdynamics2}
\end{subequations}
\noindent Here $\dot \generalcoord$ denotes the $t'$ derivative and the noise vectors are re-scaled, $\tilde{\xi}_i(t') = 2\xi_i(t)/NK_\perp$. Since we find zero damping for our specific model we also add a phenomenological damping term $\alpha$ (equal for both coordinates~\cite{Tatara2008}). In the following we drop the prime from $t'$ for brevity and consider only deterministic dynamics, neglecting the fluctuation terms $\tilde{\xi}_i$, therefore concentrating on low temperatures where fluctuations are small.

For $\massterm \ll 1$, the second derivative terms are unimportant and the resulting dynamics is the same as previously considered for DWs~\cite{Tatara2004,Tatara2008}. In particular, in the absence of the force, there is a critical spin torque $j^*_t$ above which the domain wall moves steadily. In this regime, the DW dynamics is very similar to that of superconducting Josephson junctions~\cite{Berger1986}. We discuss this correspondence further as it relates to hysteretic dynamics in Sec.~\ref{SubSec:hysteresis}.

\section{Observable Effects due to the Electron-Induced Mass\label{Sec:Experiment}}

\subsection{Resonant DW Dynamics\label{SubSec:Resonance}}
One way to explore the effect of the electron-induced domain wall mass is via resonant dynamics as in Ref.~\cite{Saitoh2004}. Resonant dynamics are investigated by applying a small amplitude ac current, which leads to an oscillating spin torque $j_{\rm t}(\omega) = j_{\rm t}e^{i\omega t}$. The oscillating spin torque causes $\chi$ and $\phi$ to oscillate as well, with the greatest amplitude oscillations occurring at the resonant frequencies of the DW. Sweeping the frequency $\omega$ and measuring the amplitude of the domain wall oscillations reveals these resonances. Typically, one measures the oscillations in position, therefore  in this section we focus on the $\chi$ response.

\begin{figure}[t!]
\centering
\includegraphics[scale=1.0]{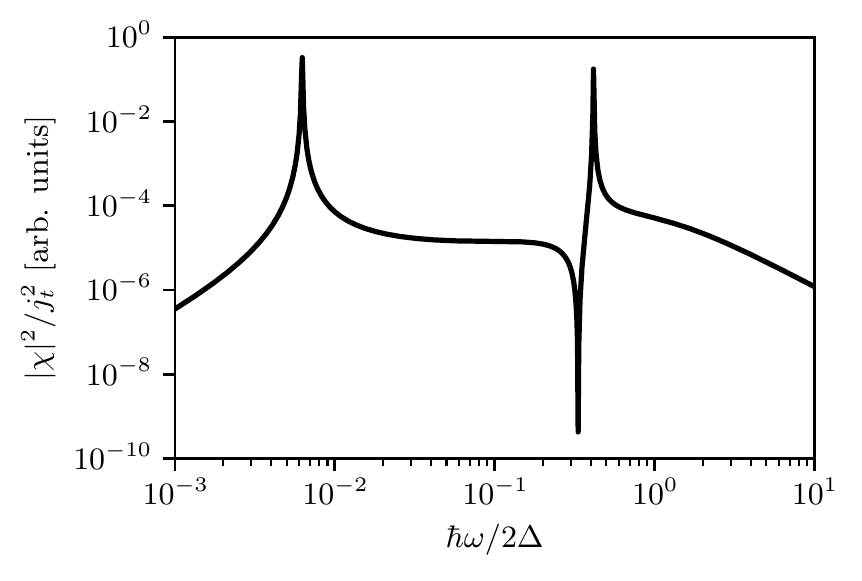}
\caption{Amplitude of DW position oscillations in response to applied ac current with amplitude $j_{\rm t}$ and frequency $\omega$. The response exhibits a low-frequency peak dependent on the pinning strength $k_{\rm p}$ and another resonance due to conduction electrons at a higher frequency. We use the parameters $s = 20$, $N=10$, $\alpha = 0.2\Delta$, $k_{\rm p} = 0.5\Delta$, $K_\perp = 8\Delta$; $\Delta$ is the strength of exchange coupling. Here we consider only the adiabatic spin torque and set $\beta_w = 0$.~\label{Fig:response}}
\end{figure}

We write the dynamical equations of motion~\eqref{Eqn:dwdynamicsgeneral} in Fourier space including the full response kernel $f^{ij}(\omega)$ from Eq.~\eqref{Eqns:fall}. In matrix form the equations are 
\begin{equation}
    A_\omega \begin{pmatrix} \phi_\omega \\ \x_\omega \end{pmatrix} = \begin{pmatrix} j_{\rm t} \\ f_x \end{pmatrix},
\end{equation}
where $j_{\rm t} \propto Pj$ is the spin torque for a current $j$ passing in a wire with spin polarization $P$~\cite{Tatara2008}. We here also include the possibility of a non-adiabatic force $f_x=\beta_w j_{\rm t}/P$ with the non-adiabaticity parameter $\beta_w$~\cite{Tatara2008}. The coefficient matrix $A_\omega$ is
\begin{equation}
    A_\omega=\begin{pmatrix} N K_\perp + f^{\phi\phi} +i\omega \alpha & i\omega N + f^{\phi \x}\\ -i \omega N + f^{\x\phi}& k_{\rm p} + f^{\x\x} + i \omega \alpha \end{pmatrix},
\end{equation}
where $k_{\rm p}$ is a spring constant from a harmonic pinning center. We include pinning in order to compare to the results in~\cite{Saitoh2004}. Here we have assumed small-amplitude dynamics of $\phi$ in order to expand $\sin(2\phi) \approx 2\phi$. It is then straightforward to numerically calculate the amplitude of oscillations in the DW position, given by $\chi_\omega^2/j_{\rm t}^2$. Figure~\ref{Fig:response} shows an example of the response for a given set of system parameters.

The response exhibits two peaks at different frequencies $\omega_{\rm p}$ and $\omega_{\rm m}$, obtained where $A^{-1}$ becomes singular. Below we present the analytical forms of $\omega_{\rm p/m}$ as $\alpha \rightarrow 0$. The first frequency $\omega_{\rm p}$ is associated with pinning and occurs at low frequencies $\omega \ll 2\Delta$. For small $k_{\rm p} \ll \Delta, K_\perp$ it is 

\begin{equation}
     \hbar\omega_{\rm p} \approx \sqrt{\frac{k_{\rm p} K_\perp \Delta}{N\Delta+K_\perp s}}.
\end{equation}

Hence for $K_\perp s \ll N\Delta$, $\omega_{\rm p}$ is independent of the parameters of the electron system matching with the result in~\cite{Saitoh2004}. This limit also corresponds to the regime where the mass parameter $m$ in Eq.~\eqref{Eqn:mass} is irrelevant, $m \ll 1$. 

The other resonance peak is associated with dynamical renormalization of the response due to the electrons. For negligible pinning ($k_{\rm p} \rightarrow 0$), 

\begin{equation}
    \hbar\omega_{\rm m} = 2\Delta \sqrt{\frac{N^2}{(N+2s)^2}+\frac{K_\perp N s}{\Delta (N+2s)^2}}.
    \label{Eqn:omegam}
\end{equation}

For $N \ll 2s$ the resonance $\omega_{\rm m}$ can be significantly lower than the frequency $2\Delta$ corresponding to the gap between the two electronic eigenstates. Figure~\ref{Fig:response} shows the response for a totally adiabatic DW ($\beta_w = 0$) for $K_\perp = 8\Delta$. In particular it shows that for $K_\perp > \Delta$ the resonance remains below the electronic spin-band gap $2\Delta$, within the bounds of our theory. We expect this resonance could be readily observed even in systems with pinning, because $\omega_{\rm p}$ and $\omega_{\rm m}$ are separated by two orders of magnitude. Increasing the damping parameter $\alpha$ merely increases the width of the response peaks and does not appreciably change the peak location.

\subsection{Hysteresis in DW Dynamics\label{SubSec:hysteresis}}

Another experimental consequence of the mass is hysteresis in the DW dynamics (not to be confused with magnetic hysteresis). Hysteresis occurs because in some regimes the equations of motion~\eqref{Eqn:dwdynamics2} have multiple solutions for the same parameter values $m$, $\alpha$, etc. The equations can yield a `running' state, with $\dot{\x}, \dot{\phi}$ approaching constant values, and a damped state where $\dot{\x}, \dot{\phi} \rightarrow 0$. This means that although there is critical torque $j^*_t$ to start the domain wall in motion, it can continue moving at a reduced  $j_{\rm t} < j^*_{\rm t}$ down to some ``retrapping" torque $j_{\rm r}$. Below we analyze this effect in more detail. 

In the absence of the non-adiabatic force and for $\massterm=0$, the equations of motion \eqref{Eqn:dwdynamics2} (without noise) can be exactly mapped to the Resistively Shunted Junction model (RSJ) model of overdamped Josephson junctions, allowing us to directly use well known results of this model~\cite{barone1982physics,terosbook}. One can first solve $\dot \x=-\dot \phi/\alpha$ and insert it into Eq.~\eqref{Eqn:dwdynamicsphi}. The equation of motion thus depends only on $\phi$. Defining $\tilde{\varphi}=2\phi$ and $\alpha'=(1+1/\alpha^2)\alpha$ yields the RSJ dynamics for the Josephson junction phase $\tilde{\varphi}$ with the damping constant $\alpha'$ and driven by the bias current $I_b = j_{\rm t} I_C$ across the junction with the critical current $I_C$. In this case the time derivative of the steady oscillating phase $\dot \phi$ of the domain wall maps to the dc voltage across the Josephson junction.

Introducing a non-vanishing inertial term $\massterm$ is analogous to the effect of capacitance for the Josephson junction problem. Here it is possible to find effects similar to those for underdamped Josephson junctions. The correspondence is no longer exact because $\x$ cannot be directly solved as a function of $\phi$, and the domain wall has more dynamical parameters than a simple Josephson junction. Comparing the Hamiltonian \eqref{Eqn:dwhamiltonian} to that of the Josephson junction suggests that the exact analogy would require considering a Cooper pair box with a dynamical external flux or gate charge~\cite{terosbook}. 

\begin{figure}[t!]
\centering
\includegraphics[scale=1.0]{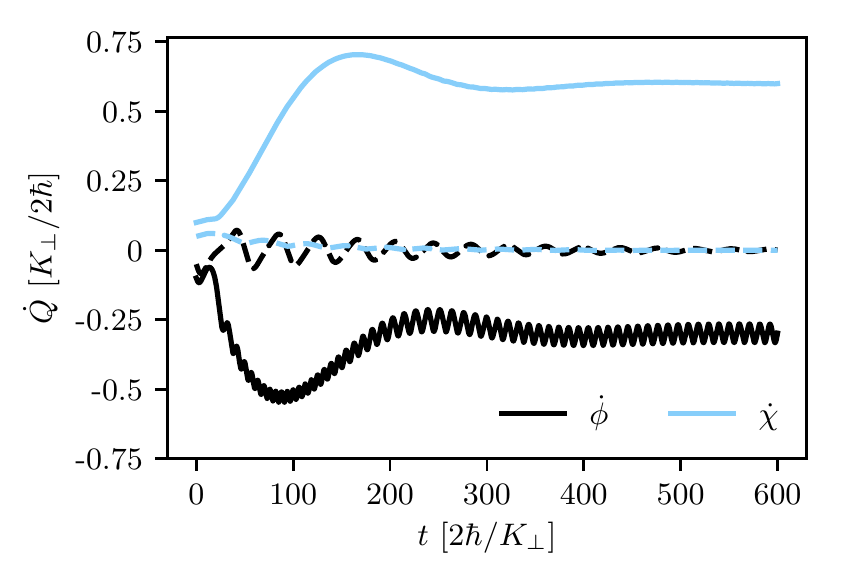}
\caption{Hysteretic dynamics of the DW as seen from two different stationary solutions of Eqs.~\eqref{Eqn:dwdynamics2}, $\dot \phi$ (black) and $\dot \x$ (blue/light gray) obtained numerically with the same system parameters $\alpha=0.5$, $\massterm=50$ and $j_{\rm t}=0.75$. The two solutions have slightly different initial conditions, leading to a running state (solid lines) and damped state (dashed lines).
\label{fig:hysteresis}}
\end{figure}

Nevertheless, we show that similar to the underdamped Josephson junctions domain walls can also exhibit hysteretic dynamics for spin torques below the critical value $j_{\rm t}^*$~\cite{barone1982physics}. In this case there is a regime of parameters $\alpha,\massterm$ and $j_{\rm t}<j^*_{\rm t}$ for which the solutions of the dynamical equations separate into different regions: (i) one with vanishing stationary values of $\dot \phi,\dot \x$ and (ii) another with non-zero time-averaged speeds. To see this, let us consider Eqs.~\eqref{Eqn:dwdynamics2} in the absence of noise terms and in the adiabatic limit where $f_x=0$. The first solution (i) corresponds to the case where $2\phi=-{\rm arcsin} (j_{\rm t})$, possible when $j_{\rm t} \le 1$ \cite{Tatara2004, Tatara2008}. The second solution (ii) can be found with the following scheme, where for simplicity we assume $\alpha \ll 1$ and $\massterm \gg 1$. We first estimate the running state values of $\dot \phi$ and $\dot \chi$ by neglecting the second derivatives and the $\sin(2\phi)$ term, which gives \begin{equation}
\begin{pmatrix} \dot \phi_0 \\ \dot \x_0 \end{pmatrix} = \frac{j_{\rm t}}{1+\alpha^2} \begin{pmatrix} -\alpha\\ 1 \end{pmatrix} \approx  \begin{pmatrix} -\alpha j_{\rm t}\\ j_{\rm t} \end{pmatrix}.
\end{equation}
We can then assume that the full dynamics is obtained from a perturbation of the running state, $\phi(t)=\phi_0+\tilde \phi$ and $\x(t) = \x_0 + \tilde \x$, where $\tilde \phi \ll \phi_0$. Ignoring $\tilde \phi$ inside the $\sin(2\phi)$ term and neglecting damping in the $\alpha \rightarrow 0$ limit, $\tilde \phi$ and $\tilde \x$ satisfy
\begin{subequations}
\begin{align}
\massterm \ddot {\tilde \phi} - \dot {\tilde \x}  + \sin\left(2 j_{\rm t} \alpha t\right) &= 0\\
\massterm \ddot {\tilde \x} + \dot {\tilde \phi}  &= 0.
\end{align}
\end{subequations}
These can be solved in closed form. Starting from the running state, we also require the initial conditions $\tilde \x(0)=\tilde \phi(0)=\dot{\tilde \x}(0)=\dot{\tilde \phi}(0)=0$. We get 
\begin{align}
\tilde \x(t) &= \frac{\cos(2 j_{\rm t} \alpha t)-1+4 j_{\rm t}^2 \massterm^2 \alpha^2[1-\cos(t/\massterm)]}{8 j_{\rm t}^3 m^2 \alpha^3 - 2 j_{\rm t} \alpha}  \\
\tilde \phi(t) &= \frac{\massterm \left[\sin(2 j_{\rm t} \alpha t)-2 j_{\rm t} \massterm \alpha \sin(t/\massterm)\right]}{4 j_{\rm t}^2 \massterm^2 \alpha^2-1}. 
\label{eq:phitilde}
\end{align}
From these, only the magnitude of $\tilde \phi(t)$ constrains the approximation above. The deviation from the running state is small if $\max_t |\dot {\tilde \phi}| \ll  \dot \phi_0$. Based on Eq.~\eqref{eq:phitilde}, this can happen in two cases depending on the magnitude of $2 j_{\rm t} \massterm \alpha$. If $2 j_{\rm t} \massterm \alpha$ is small, the requirement of a small deviation from the running state leads to $m \ll 1$, which is inconsistent with the assumptions made above. Therefore we must assume $2 j_{\rm t} \massterm \alpha \gg 1$, and  we can neglect the first term in the numerator and the second term in the denominator of Eq.~\eqref{eq:phitilde}. The condition of a small $\dot{\tilde{\phi}}$ then yields $1/2 j_{\rm t} m \alpha \ll j_{\rm t} \alpha$, or in other words $j_{\rm t} > j_{\rm r}$ with a retrapping torque (in analogy to the Josephson junction retrapping current~\cite{barone1982physics}) 
\begin{equation}
    j_{\rm r} = \frac{c}{\alpha \sqrt{2 \massterm}},
    \label{eq:retrappingtorque}
\end{equation}
where $c$ is a number of the order of unity. From the $j_{\rm r}$ obtained by numerically solving the full dynamical equations we find $c \approx 2$. The rather counterintuitive increasing of the hysteresis (i.e., decreasing $j_{\rm r}$) upon increasing $\alpha$ is limited to small $\alpha$. When $\alpha$ becomes of the order of unity, the trend is reversed and $j_{\rm r}$ increases with increasing $\alpha$. From Eq.~\eqref{eq:retrappingtorque} we also get a requirement for $\massterm$ since hysteresis takes place only if the two solutions (i) and (ii) coexist, i.e., $j_{\rm r} < 1$. This yields a condition for the mass term allowing hysteretic dynamics,
\begin{equation}
    \massterm \gtrsim \massterm_{\rm h} \equiv \frac{2}{\alpha^2}.
\end{equation}
We confirm this analysis in one case by showing the numerically simulated dynamics of $\phi$ and $\x$ in Fig.~\ref{fig:hysteresis} for fixed system parameters $\alpha, m,$ and $j_{\rm t}$.

\begin{figure}[t!]
\centering
\includegraphics[scale=1.0]{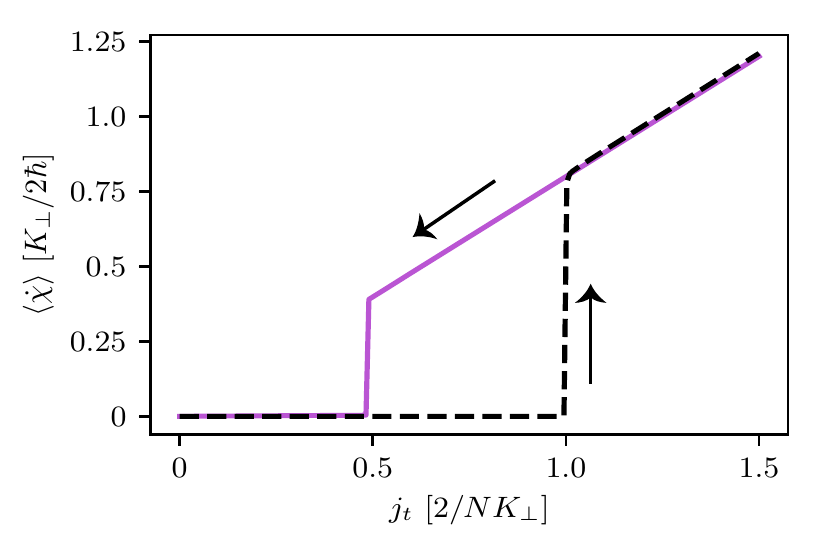}
\caption{Hysteresis loop for DW dynamics. For a range of spin torques, the time-averaged DW speed can have multiple values depending on whether the spin torque is increased (black dashed curve) or decreased (solid purple/light gray curve). Here we numerically simulated Eqs.~\eqref{Eqn:dwdynamics2} with different initial conditions for $m=50$ and $\alpha=0.5$.\label{fig:hyst}}
\end{figure}

For $\massterm > \massterm_{\rm h}$ and in the absence of the non-adiabatic force, the domain wall can be set to a fixed speed only once the spin torque exceeds the critical value $j^*_{\rm t}$. However, when reducing $j_{\rm t}$ below the critical torque the DW can stay in the running state until the motion gets trapped into a minimum of the washboard potential $V(\phi,\x)$ for spin torques below the retrapping torque $j_{\rm r}$. In Fig.~\ref{fig:hyst} we show an example of a hysteresis loop for the time-averaged domain wall speed, $\langle \dot \x \rangle$. If the domain wall is already in the running state (upper branch), it continues to move even as the torque is reduced below the critical torque. The retrapping torque in our numerical simulation is $j_{\rm r} \approx 0.5$ whereas the critical torque is $j^*_t = 1$. Further exploration of the stability of the running state for different parameters will be discussed elsewhere.

One possible way to reveal the hysteretic dynamics of domain walls is to perform experiments with pulsed currents, similar to Ref.~\cite{Thomas2010}, but with currents leading to torques close to the critical torque $j_{\rm t}^*$. In~\cite{Thomas2010} it was possible to study the distance spanned by a domain wall within a given time after an initial current pulse, including deceleration after the pulse was switched off. Let us consider the case where an experiment uses two pulses of different heights, corresponding to torques $j_1$ and $j_2$ applied on the domain wall. The current pulses should be chosen such that $j_{\rm r} < j_1 < j_{\rm t}^*$ and $j_{\rm t}^*<j_2$. This way, one of the torques is above the critical torque, whereas one is between the retrapping torque $j_{\rm r}$ and the critical torque.

In this case the distance traversed by the domain wall depends on the order of these current pulses. In the first experiment, $j_1$ is applied first and the domain wall goes to the running state only after applying $j_2$. The second experiment corresponds to the opposite case where $j_2$ is applied first and induces steady motion, and then applying $j_1$ maintains the DW motion because of hysteresis. Therefore, the distance travelled by the domain wall as a result of the current pulses is larger in the second experiment. 

\section{Conclusions\label{Sec:Discussion}}

This paper is devoted to the analysis of domain wall dynamics in a metallic ferromagnetic nanowire. The key finding of this work is that coupling of the magnetic texture to conduction electrons gives rise to the DW's effective mass and two-component Langevin noise in the equations of motion. These DW equations of motion represent a new type of dynamical system with a rich variety of dynamical behaviors. We specifically discuss two examples of such novel dynamical phenomena: resonant dynamics of the DW in response to an ac current and its hysteretic motion. 

From a broader perspective, our work belongs to a long list of studies discussing the origin of effective mass of topological textures in the order parameter in various ordered quantum phases (e.g., vortices in superconductors and superfluids, domain walls, vortices, and skyrmions in magnets, etc). The fundamental question of the origin and value of the effective mass of such defects has been  controversial (for example, there are conflicting statements about the mass of a superfluid vortex discussed in the literature, see~\cite{Kopnin1978, Nikolic2007, Thouless2007, Simula2018} and references therein). Likewise, domain wall or similarly soliton dynamics in magnets and superfluids have proven to be a  non-trivial problem due to the integrable structure of the theory~\cite{Braun1996, Efimkin2016, Psaroudaki2017, Kim2018}. However, most of these complications are due to the choice of model, where the effective mass and dissipation are sought to arise ``internally'' from the coupling of the defects in the order parameter field to low-energy excitations (spin waves, phonons, Bogoliubov excitations, etc) in the same field. 

While this represents an interesting and challenging theoretical problem, the simple observation put forward in this paper is that in most solid-state systems, there are external baths (e.g., phonons of the underlying crystal lattice and/or itinerant electrons in metallic systems, as explicitly considered here) that provide an alternative mechanism for an effective mass to arise. This external origin of the effective mass is non-universal, but can be dominant in some actual material systems. Since these excitations in the bath represent a system different from the fluctuations in the order parameter field itself, scattering of those off of the defect is free of complications due to self-consistency and possibly integrability constraints.  Therefore, the corresponding theoretical description is simpler than the self-consistent treatment required for the problem of the effective mass of internal origin.

This paper developed a general theoretical framework to describe dynamics of rigid magnetic textures in the presence of conduction electrons in ferromagnets. The appropriate method involves a combination of the Keldysh technique and collective coordinate approach that has been introduced originally to describe solitons in field theories~\cite{Rajaraman1987}. Minimizing the Keldysh action gives rise to the quasiclassical equations of motion that govern real-time dynamics of the defect in response to both external torques and stochastic Langevin forces. The two-component equations involve a correlated ``matrix'' noise  and a matrix response kernel that contains both dissipative effects and a contribution to the effective mass. We also formulate a generalized fluctuation-dissipation theorem. We apply the Keldysh collective-coordinate method specifically to the case of a planar domain wall in a ballistic (quasi)-one-dimensional ferromagnetic wire. The domain wall is described by two coordinates: the actual position of the domain wall and its tilt-angle. An effective mass is then shown to arise for both coordinates renormalizing the DW equations of motion. 

Here we estimate this effective mass $m$ for Co/Ni nanowires, a common material for domain wall experiments~\cite{Tanigawa2009, Koyama2011}. Assuming a quadratic dispersion for electrons, recall that $m \approx K_\perp\lambda/Nhv_{\rm F}$. We estimate $K_\perp \approx K_{\rm u}\lambda w^2$ where $K_{\rm u}$ is the reported uniaxial perpendicular anisotropy, $\lambda$ is the domain wall width and $w$ is the transverse width of the nanowire. For CoNi nanowires we use $v_{\rm F} \sim 10^6~\mathrm{m/s}$ and $K_{\rm u}\approx 5\times10^5~\mathrm{J}/\mathrm{m}^{3}$~\cite{Tanigawa2009, Koyama2011}. We find $m \approx 9$ for a domain wall with $\lambda \approx 100~\mathrm{nm}$, $w \approx 50~\mathrm{nm}$ and lattice constant $a\sim~1~$\AA. This shows that in conventional materials with high Fermi velocity, $m$ can be $\sim\mathcal{O}(10)$. Therefore $m$ could be relevant for domain wall dynamics, 
particularly in the transient regime.

We also estimate the resonant frequency $\omega_{\rm m}$ in Eq.~\eqref{Eqn:omegam} due to coupling to electrons. Measuring the value of the resonant frequency can be viewed as a proxy for measuring the inertial mass (the lower the mass, the higher the resonance frequency). Estimating $s\sim N$ and the exchange coupling $\Delta\sim 0.1~\mathrm{eV}$~\cite{Tatara2008}, we find $\omega_{\rm m}\sim 10^4~\mathrm{THz}$. Thus, in these materials the resonant frequency is probably too high to be observed. However, for materials with smaller $K_\perp$ it may be visible with current experimental techniques. It is likely that in most materials $\omega_{\rm m}$ will be higher than any pinning-induced resonances, which were reported to be in the MHz range in Ref.~\cite{Saitoh2004}.

At a qualitative level, the value of the electron-induced mass, we found, can be related to electron time-of-flight through the defect. This may represent a generic qualitative result valid more broadly than the specific problem we study. For standard ballistic metals the corresponding time-scale is generally small, but we argue that there still exist observable phenomena associated with the emergent DW mass (e.g., the hysteretic dynamics discussed in Sec.~\ref{SubSec:hysteresis} should be observable in garden-variety metallic systems, while the resonance discussed in Sec.~\ref{SubSec:Resonance} is in the Terahertz regime). 

An interesting question is how to enhance the value of the mass. This is also important for applications, since a larger mass corresponds to a wider hysteresis loop and therefore a lower driving current needed to sustain domain wall motion. As far as ballistic systems go, candidate materials include flat band systems and in general materials with itinerant electron bands' having a smaller effective Fermi velocity and correspondingly longer time-of-flight scales. The inertial DW mass, $m$, can  also be made much larger in the case of strong perpendicular anisotropy.  

Another, perhaps more experimentally relevant observation is that disorder should enhance the electron-induced effective mass. Estimates of the diffusion time scale through the DW in typical disordered ferromagnets  give rise to an enhanced value of the mass  and a resonance frequency much smaller than expected for ballistic metals. Specifically, the inertial effective mass induced by diffusive  electrons with the mean free path $\ell$ is expected to be $\lambda/ \ell$ times larger than for the same domain wall of width $\lambda$ in a ballistic metal. For a typical nanowire, the mean free path is around $\ell \sim 10~\mathrm{nm}$, which implies about a ten-fold enhancement of the mass. We expect that the enhancement can be larger in more disordered systems, but a more accurate analysis is required to precisely determine domain wall dynamics in diffusive metals.  A detailed theory of domain wall dynamics in disordered ferromagnets will be presented elsewhere.

In the presence of the fluctuation terms $\xi_\phi(t)$ and $\xi_\x(t)$, we expect the domain wall to behave qualitatively similarly to the Josephson junction affected by noise: In particular, when temperature is not much smaller than $K_\perp$, the thermal fluctuations lead to a ``premature switching'' to the finite-speed state even for $j_{\rm t} < j^*_{\rm t}$ \cite{Ivanchenko1969josephson}. Another intriguing possibility would be to study the dynamics of the stochastic ``escape'' process of the domain wall from its potential minimum and investigate if this system could be driven to the regime of macroscopic quantum tunneling as in the case for Josephson junctions~\cite{Takagi1996, Voss1981,Devoret1985}. This would correspond to studying the statistics of domain wall motion with pulsed currents, in an experiment analogous to \cite{Thomas2010}. Such experiments would also be useful in determining the error rates of magnetic texture-based magnetic memories. 

\section*{Acknowledgements}
We are grateful to Rembert Duine, Johannes Hofmann, Daniel Loss, Christina Psaroudaki, Gen Tatara, Yaroslav Tserkovnyak, and Pauli Virtanen for useful discussions.  H.M.H. acknowledges the support of an NRC Research Assistantship at NIST. V.G. was supported by DOE-BES (DESC0001911) and the Simons Foundation. The work of T.T.H. was supported by the Academy of Finland (project number 317118). T.T.H. is also grateful to the Physics Frontier Center at the JQI, where this work was conceived, for hospitality. 
\appendix
\section{Exact Response Kernel Expressions \label{App:response}}
The fermionic Keldysh-space Green's function matrix is 
\begin{equation}
\check{G}_{\sigma k}(t,t') = \begin{pmatrix}
G^{\rm R}_{\sigma k}(t,t') & G^{\rm K}_{\sigma k}(t,t') \\
0 & G^{\rm A}_{\sigma k}(t,t'), 
\end{pmatrix}
\end{equation}
where 
\begin{subequations}
\begin{align}
G^{\rm R}_{\sigma k}(t,t') &= -i\Theta(t-t')e^{-i\varepsilon_k^\sigma(t-t')}\\
G^{\rm A}_{\sigma k}(t,t') &= ~~i\Theta(t'-t)e^{-i\varepsilon_k^\sigma(t-t')}\\
G^{\rm K}_{\sigma k}(t,t') &= -i(1-2f_{\sigma k})e^{-i\varepsilon_k^\sigma(t-t')}
\end{align}
\label{Eqn:FreeGFs}
\end{subequations}
\noindent \noindent are the retarded (R), advanced (A), and Keldysh (K) parts of the Green's function and $\Theta(t-t')$ is the Heaviside function. Assuming the electrons to be in thermal equilibrium, $f_{\sigma k} = \left(e^{\varepsilon_{\sigma k}-\mu/T}+1\right)^{-1}$ is the Fermi-Dirac distribution. The response kernel and noise correlation function are defined in terms of the retarded (R) and Keldysh (K) polarization functions $\Pi_{ij}$,  
\begin{align}
    \eta^{ij}(t,t') &= \frac{\partial\Pi^{\rm R}_{ij}(t,t')}{\partial t}, \\
    \mathcal{C}^{ij}(t,t') &= -\frac{i}{2}\frac{\partial^2\Pi^{\rm K}_{ij}(t,t')}{\partial t \partial t'},
\end{align}
where $\Pi_{ij}$ has a matrix structure due to the two collective coordinate perturbations. The polarization functions are 
\begin{widetext}
\begin{align}
\Pi^{\rm R}_{ij}(t,t') &= i\Theta(t-t')\sum_{\substack{\sigma\sigma' \\k k'}}~^i\!V^{\sigma\sigma'}_{kk'}~^j\!V^{\sigma'\sigma}_{k'k}\left[f_{\sigma k} - f_{\sigma' k'}\right]e^{-i(\varepsilon^{\sigma'}_{k'}-\varepsilon^{\sigma}_{k})(t-t')}\label{Eqn:PiR}, \\
\Pi^{\rm K}_{ij}(t,t') &= i\sum_{\substack{\sigma\sigma' \\k k'}}~^i\!V^{\sigma\sigma'}_{kk'}~^j\!V^{\sigma'\sigma}_{k'k}\left[f_{\sigma' k'} + f_{\sigma k} - 2f_{\sigma'k'} f_{\sigma k}\right]e^{-i(\varepsilon^{\sigma'}_{k'}-\varepsilon^{\sigma}_{k})(t-t')}.\label{Eqn:PiK}
\end{align}
Therefore, 
\begin{align}
\eta^{ij}(t-t') &= \frac{i}{2}\delta(t-t')\sum_{\substack{\sigma\sigma' \\k k'}}~^i\!V^{\sigma\sigma'}_{kk'}~^j\!V^{\sigma'\sigma}_{k'k}\left[h_{\sigma'k'}-h_{\sigma k}\right]e^{-i(\varepsilon_{\sigma' k'}-\varepsilon_{\sigma k})(t-t')}\nonumber \\ 
&+ \frac{1}{2}\Theta(t-t')\sum_{\substack{\sigma\sigma' \\k k'}}~^i\!V^{\sigma\sigma'}_{kk'}~^j\!V^{\sigma'\sigma}_{k'k}\left[h_{\sigma'k'}-h_{\sigma k}\right](\varepsilon_{\sigma' k'}-\varepsilon_{\sigma k})e^{-i(\varepsilon_{\sigma' k'}-\varepsilon_{\sigma k})(t-t')}\\
C^{ij}(t-t') &= \frac{1}{4}\sum_{\substack{\sigma\sigma' \\k k'}}~^i\!V^{\sigma\sigma'}_{kk'}~^j\!V^{\sigma'\sigma}_{k'k}\left[1-h_{\sigma k}h_{\sigma' k'}\right](\varepsilon_{\sigma' k'}-\varepsilon_{\sigma k})^2e^{-i(\varepsilon_{\sigma' k'}-\varepsilon_{\sigma k})(t-t')}
\end{align}
\end{widetext}
where $h_{\sigma k}  = \tanh(\varepsilon_{\sigma k}-\mu/2T)$ is used in place of the Fermi-Dirac distribution $f_{\sigma k}$, since $h_{\sigma k} = 1 - 2 f_{\sigma k}$. The FDT in the main text can be derived from these expressions. 
\section{Eigenstates for Adiabatic Domain Wall \label{App:eigenstates}}
The single particle Schr\"{o}dinger equation including the static domain wall from Eq.~\eqref{Eqn:DWpotential} is the eigenvalue equation $H[\bs{\Omega}_0(x)]\bs{\varphi}_k(x) = \varepsilon\bs{\varphi}_k(x)$, where $\bs{\varphi}_k$ is a two-component spinor wavefunction and $\varepsilon$ is an eigenvalue giving the energy of the state with wavefunction $\bs{\varphi}_k(x)$. Recall that for the static domain wall solution we have 
\begin{equation}
  \Delta\bs{\Omega}_0(\lambdaratio x)\cdot\hat{\bs{\tau}} = \Delta\tanh\left(\lambdaratio x\right)\hat{\tau}^1 + \Delta\sech\left(\lambdaratio x \right)\hat{\tau}^3,  
  \label{AppEqn:DWpotential}
\end{equation}
where $\lambdaratio = \lambda_{\rm F}/\lambda$ and $x$ is dimensionless. Thus, we seek solutions to the equation $\hat{D}\bs{\varphi} = 0$ where
\begin{equation}
\hat{D} = 
    \begin{pmatrix}
    \hat{h}_0 - \Delta\tanh(\lambdaratio x) -\varepsilon & -\Delta\sech(\lambdaratio x) \\ 
    -\Delta\sech(\lambdaratio x) & \hat{h}_0 + \Delta\tanh(\lambdaratio x) - \varepsilon
    \end{pmatrix}, 
    \label{Eqn:Schrodinger}
\end{equation}
and $\hat{h}_0 = \hbar^2\hat{k}^2/2m_{\rm e}\lambda_{\rm F}^2$ with $\hat{k} = -i\partial_x$. Since we assume $\lambdaratio \ll 1$, the domain wall texture is a slowly-varying potential for the electrons and we can use semiclassical methods to find the eigenstates $\bs{\varphi}_k(x)$. Here we employ a method using the Weyl correspondence and an expansion in $\hbar$ which was developed in Ref.~\cite{Littlejohn1991} to treat multi-component wave equations. We briefly outline the method and refer the interested reader to~\cite{Littlejohn1991} for more details. 

For separable matrix operators $\hat{D}(\hat{x}, \hat{k}) = \hat{D}^1(\hat{x}) + \hat{D}^2(\hat{k})$, the Weyl correspondence is simply $\hat{x}\rightarrow x$ and $\hat{k}\rightarrow k$. The matrix $\hat{D}(x, k)$ solves the eigenvalue equation
\begin{equation}
    \hat{D}(x, k)\bs{\psi}^{(\sigma)}(x, k) = \epsilon^{(\sigma)}(x, k)\bs{\psi}^{(\sigma)}(x, k), 
\end{equation}
with eigenvectors $\bs{\psi}^{(\sigma)}(x, k)$ and eigenvalues $\epsilon^{(\sigma)}(x, k)$, where $\sigma=1,2$ is an index labeling each eigenvalue and corresponding eignevector. The original wavefunctions $\bs{\varphi}$ can be written as
\begin{equation}
    \bs{\varphi}^{(\sigma)}_k(x) = \bs{\psi}^{(\sigma)}_k(x)\tilde{\psi}^{(\sigma)}(k, x)
    \label{Eqn:wkbansatz}
\end{equation}
where $\tilde{\psi}$ is \emph{not} a spinor and solves the single-component wave equation
\begin{equation}
    \epsilon^{(\sigma)}(\hat{x}, \hat{k})\tilde{\psi}^{(\sigma)} = 0
\end{equation}
where the operators $\hat{x}, \hat{k}$ are restored. The first order correction to $\tilde{\psi}$ is found by solving a new wave equation $\epsilon^{(\sigma),1}(\hat{x}, \hat{k})\tilde{\psi}^{(\sigma),1} = 0$ where the first order correction to $\epsilon$ is~\cite{Littlejohn1991}
\begin{align}
    \epsilon^{(\sigma),1} &= -i\bs{\psi}^{\dagger(\sigma)}\left\lbrace \bs{\psi}^{(\sigma)}, \epsilon^{(\sigma)}_0\right\rbrace \nonumber \\
    &-\frac{i}{2}\left(\hat{D}-\epsilon^{(\sigma)}_0\mathbb{I}\right) \left\lbrace \bs{\psi}^{\dagger(\sigma)}, \bs{\psi}^{(\sigma)}\right\rbrace.
    \label{Eqn:correction}
\end{align}
The notation $\left\lbrace \cdot \right\rbrace$ is a classical Poisson bracket and $\epsilon_0^{(\sigma)}$ is the zeroth-order eigenvalue. For the domain wall, $\hat{D}$ has normalized eigenvectors
\begin{subequations}
\begin{align}
    \bs{\psi}^{(1)}_k(x) &= \frac{1}{\sqrt{1+e^{-2\lambdaratio x}}}\begin{pmatrix}-e^{-\lambdaratio x}\\1\end{pmatrix}, \\
    \bs{\psi}^{(2)}_k(x) &= \frac{1}{\sqrt{1+e^{-2\lambdaratio x}}}\begin{pmatrix}1\\ e^{-\lambdaratio x}\end{pmatrix}.
\end{align}
\end{subequations}
The corresponding eigenvalues are
\begin{subequations}
\begin{align}
    \epsilon^{(1)}(\hat{k}) &= \frac{\hbar^2\hat{k}^2}{2m_{\rm e}\lambda_{\rm F}^2} - \mu + \Delta -\varepsilon, \\
    \epsilon^{(2)}(\hat{k}) &= \frac{\hbar^2\hat{k}^2}{2m_{\rm e}\lambda_{\rm F}^2} - \mu - \Delta -\varepsilon.
\end{align}
\label{Eqn:eigvals}
\end{subequations}
The solutions to the wave equations using eigenvalues~\eqref{Eqn:eigvals} are simply plane waves, $\tilde{\psi}^{(\sigma)}_k(x) = e^{ikx}$. The first-order correction using Eq.~\eqref{Eqn:correction} is zero for this case. Hence, using the ansatz \eqref{Eqn:wkbansatz} gives the final solutions used in the main text, 
\begin{subequations}
\begin{align}
    \bs{\varphi}^{(1)}_{k}(\lambdaratio x) &= \frac{1}{\sqrt{1+e^{-2\lambdaratio x}}}
    \begin{pmatrix}
    -e^{-\lambdaratio x} \\
    1
    \end{pmatrix}
    e^{ikx}
    \label{appEqn:psiup}
    \\
    \bs{\varphi}^{(2)}_{k}(\lambdaratio x) &= \frac{1}{\sqrt{1+e^{-2\lambdaratio x}}}
    \begin{pmatrix}
    1\\
    ~~e^{-\lambdaratio x}
    \end{pmatrix}
    e^{ikx}.
    \label{appEqn:psidown}
\end{align}
\label{AppEqn:basis}
\end{subequations}
In the main part of the paper we use the labels $\uparrow, \downarrow$ for the eigenstates instead of $1,2$. This is to emphasize the aligned and anti-aligned solutions for the static domain wall. One can check that Eqs.~\eqref{AppEqn:basis} form a complete basis, where 
\begin{equation}
    \int dx~\bs{\varphi}^{\dagger(\sigma)}_k(x)\bs{\varphi}^{(\sigma')}_{k'}(x) = 2\pi\delta_{\sigma\sigma'}\delta_{kk'}, 
\end{equation}
and 
\begin{equation}
    \sum_{\sigma}\int~dkdx~\bs{\varphi}^{\dagger(\sigma)}_k(x)\bs{\varphi}^{(\sigma)}_{k}(x) = 1.
\end{equation}

\section{Calculation of $f^{ij}(\omega)$ \label{App:F}}
Here we present additional details of the calculation for the inertial part of the response function. The expression for $f^{ij}(\omega)$ can be further simplified by studying the symmetry properties of the matrix elements $^i\!V^{\sigma\sigma'}_{kk'}$ for the diagonal and off-diagonal cases. The diagonal components are 
\begin{equation}
    f^{ii}(\omega) = \frac{\hbar^2\omega^2}{2}\sum_{\substack{\sigma\neq\sigma'\\K q}}~\left|^i\!V^{\sigma\sigma'}_{q}\right|^2\left[\frac{h_{\sigma'K^-}-h_{\sigma K^+}}{\hbar\omega - (\varepsilon_{\sigma' K^-}-\varepsilon_{\sigma K^+})}\right].
\end{equation}
where $K = (k+k')/2$, $q = k-k'$, and $K^\pm = K\pm q/2$. The intraband matrix elements $\left|^i\!V^{\sigma\sigma}_{q}\right|^2 \approx \delta(q)$, and therefore their contribution is zero. We can now explicitly write the sum over $\sigma$, which gives
\begin{widetext}
\begin{align}
    f^{ii}(\omega) &= \frac{\hbar^2\omega^2}{2}\int \frac{dKdq}{(2\pi)^2}~\left|^i\!V^{\downarrow\uparrow}_{q}\right|^2\left[\frac{h_{\uparrow K^-}-h_{\downarrow K^+}}{\hbar\omega - (\varepsilon_{\uparrow K^-}-\varepsilon_{\downarrow K^+})} + \frac{h_{\downarrow K^-}-h_{\uparrow K^+}}{\hbar\omega - (\varepsilon_{\downarrow K^-}-\varepsilon_{\uparrow K^+})}\right] \nonumber \\
    &\approx \frac{4\hbar^2\omega^2\Delta}{(\hbar\omega)^2-4\Delta^2}\frac{1}{4\lambdaratio}\int \frac{dK}{2\pi}(h_{\uparrow K}-h_{\downarrow K}) \nonumber \\
    &\approx \frac{4\Delta\hbar^2\omega^2}{(\hbar\omega)^2-4\Delta^2}s
    \label{Eqn:fgammagamma}
\end{align}
\end{widetext}
The interband matrix elements $~|^i\!V^{\downarrow\uparrow}_{q}|^2 \propto \sech^2(\pi q/2\lambdaratio)$ strongly suppress scattering for high values of $q$, and so we assume that $Kq \sim k_{\rm F}q \ll \Delta$ and do not consider these terms. Finally, in the last line we do the integral over $K$ by taking the zero temperature limit $h_{\sigma K} \approx \sgn(\varepsilon_{\sigma K})$.  

The cross term $^i\!V^{\sigma\sigma'}_{kk'}~^j\!V^{\sigma'\sigma}_{k'k}$ is antisymmetric upon exchanging $\sigma \leftrightarrow \sigma'$ and $k \leftrightarrow k'$. We employ the same approximations used in Eq.~\eqref{Eqn:fgammagamma} and find the following
\begin{widetext}
\begin{align}
    f^{ij}(\omega) &= \frac{\hbar^2\omega^2}{2}\sum_{Kq}~^i\!V^{\uparrow\downarrow}_q~^j\!V^{\downarrow\uparrow}_{q}\left[\frac{h_{\downarrow K^-}-h_{\uparrow K^+}}{\hbar\omega - (\varepsilon_{\downarrow K^-}-\varepsilon_{\uparrow K^+})} - \frac{h_{\uparrow K^-}-h_{\downarrow K^+}}{\hbar\omega - (\varepsilon_{\uparrow K^-}-\varepsilon_{\downarrow K^+})}\right]\nonumber \\
    &\approx \pm \frac{2i\hbar^3\omega^3}{(\hbar\omega)^2-4\Delta^2}\frac{1}{4\lambdaratio}\int\frac{dK}{2\pi}(h_{\uparrow K} - h_{\downarrow K})\nonumber\\
    &\approx \pm\frac{2i\hbar^3\omega^3s}{(\hbar\omega)^2-4\Delta^2}.
\end{align}
\end{widetext}
\bibliography{main}
\end{document}